\documentstyle[aps,amsmath,epsfig]{revtex}

\begin{document}
\title{The Kraichnan-Kazantsev dynamo}
\author{D.~Vincenzi}
\address{CNRS, Observatoire de la C\^ote d'Azur, B.P. 4229,
06304 Nice Cedex 4, France\\
Dipartimento di Fisica, Universit\`a di Genova, Via Dodecaneso, 33, 
I-16142, Genova, Italy}
\date{\today}
\begin{titlepage}
\maketitle
\vspace{3cm}
\noindent
Suggested running head: The Kraichnan-Kazantsev dynamo\\
\vfill
\begin{list}{}{\leftmargin 11em \labelsep 1em
\labelwidth \leftmargin \advance\labelwidth-\labelsep
\topsep 0.1em \parsep 0.5pt plus 0.25pt minus 0.25pt \itemsep \parsep}
\item[Mailing Address:\hfill]
Dario Vincenzi\\
CNRS UMR 6529\\
Bd. de l'Observatoire\\
BP 4229\\
06304 NICE CEDEX 4 - France
\item[Telephone number:\hfill]+33-4-92003172
\item[Fax number:\hfill]+33-4-92003121
\item[E-mail address:\hfill] vincenzi@obs-nice.fr
\end{list}
\end{titlepage}
\begin{abstract}
The problem of the dynamo effect for a Kraichnan 
incompressible helicity-free
velocity 
field is considered. 
Exploiting a quantum formalism first introduced by 
Kazantsev (A.P.~Kazantsev, Sov. Phys. JETP {\bf 26}, 1031-1034 (1968)), 
we show that a 
critical magnetic Reynolds number exists for the presence 
of dynamo. 
The value of the Prandtl number 
influences the spatial distribution
of the magnetic field and its growth in time.
The magnetic field correlation length is always the largest between the
diffusive scale and the viscous scale of the flow.
In the same way the field growth is characterized 
by a time scale 
that corresponds to the largest between the diffusive 
and the viscous characteristic time.
\end{abstract}

\section{Introduction}

The study of magnetic fields generated by the turbulent motion 
of a charged
conducting fluid is relevant for several astrophysical applications
\cite{Moffatt}.
An inhomogeneous flow of a charged fluid is able to locally
produce a magnetic field. The advected field  
in turn generates electric currents in the fluid and these dissipate the magnetic energy
because of the finite resistivity.
Depending on the properties of the flow, magnetic field creation or 
dissipation can prevail.\\ 
To determine in general terms the presence or not of dynamo is a
daunting task. However, as we will see, there are particular models which allow
for a detailed treatment.

The evolution of an initially given magnetic field
${\mathbf B}({\mathbf r},0)$ in an incompressible 
flow of a conducting fluid is determined by the following
equations \cite{Moffatt}
\begin{equation}
\label{B}
\begin{cases} 
\partial_t{\mathbf B}+({\mathbf v}\cdot\nabla){\mathbf B}=
({\mathbf B}\cdot\nabla){\mathbf v}+\kappa\nabla^2{\mathbf B}\\
\nabla\cdot{\mathbf B}=0
\end{cases}
\end{equation}
where ${\mathbf v}({\mathbf r},t)$ is the velocity field.
The magnetic diffusivity $\kappa$, which is assumed to be uniform 
and constant, is
proportional to the inverse of the electric conductivity
of the fluid.\\
In Eqs. \eqref{B} the term 
$({\mathbf v}\cdot\nabla){\mathbf B}$ is a purely advective
contribution that preserves the magnetic energy. The stretching term
$({\mathbf B}\cdot\nabla){\mathbf v}$  
acts either as a energy source or as a sink  
depending on the local properties of the flow. 
Finally,
the diffusive term $\kappa\nabla^2{\mathbf B}$ is responsible
for the ohmic dissipation at small scales
and balances with the inertial terms at the diffusive scale $r_d$.\\
The relative importance of the two contributions on the right-hand side
of \eqref{B} is given by the magnetic Reynolds number $Re_m
=UL/\kappa$, where $L$ denotes the integral scale of the 
flow
and $U$ is the characteristic velocity at such scale.
$Re_m$ can be regarded as a dimensionless measure 
of the fluid conductivity. 
For $Re_m\to 0$ the diffusion dominates
and the magnetic energy density (proportional to $B^2$) always 
decays to zero in time. In the opposite limit, $Re_m\to \infty$,
the diffusion term is relevant only at very small scales
and the magnetic
field is almost frozen in the fluid. 
We can expect that at high magnetic Reynolds numbers the
flow is able to enhance the magnetic field, producing
a consequent growth in time of $B^2$.
The last process is 
called
dynamo effect, referring to the energy transfer from the 
velocity field to the magnetic one.\\
The field $\mathbf B$ acts on the velocity by means of the  
Lorentz force, which yields a term proportional to
$({\mathbf B}\cdot\nabla){\mathbf B}$ in Navier-Stokes equations.
In general, it would be necessary to take into account such
feedback action on $\mathbf v$.
However, since we are interested in 
understanding if
the initial generation 
of the magnetic field is 
a persistent situation or not, 
we can assume for the initial conditions
$B^2\ll v^2$ and neglect the Lorentz
force contribution.
Under this hypothesis the evolution equations \eqref{B} are
totally uncoupled from Navier-Stokes equations. 
Following this kinematic approach, we thus proceed
as if $\mathbf v $ was an assigned random field: 
given the initial condition
${\mathbf B}({\mathbf r},0)$
and appropriate boundary conditions, Eqs. \eqref{B} 
completely
determine the magnetic field evolution.

For the prescribed velocity we refer to the Kraichnan 
statistical ensemble
\cite{Kraich}, in which $\mathbf v$ is assumed Gaussian,
homogeneous, isotropic and $\delta$-correlated in time.
The reason for this choice is that  analytical results 
can be obtained \cite{Kazantsev,V96}.
A real turbulent flow is characterized by two scales:
the integral scale $L$ and the viscous scale $\eta$, at which
the dissipation term and the transport one balance in Navier-Stokes
equations.
The velocity structure 
function is supposed to be smooth for $r\ll \eta$,
to scale as $r^\xi$ $(0\leq \xi
\leq 2)$ in the inertial range $\eta\ll r\ll L$
and to approach a constant value at scales much larger than $L$.
The parameter $\xi$ 
represents the H\"older exponent of the structure function and 
can be thought of as a measure of the field roughness: for $\xi=2$ the velocity
is smooth in space, while with $\xi=0$ we describe a diffusive field.
It is well known that the magnetic dynamo can emerge for an helical 
flow due to the $\alpha$-effect \cite{Moffatt,Krause}. 
Here we will restrict to a parity invariant
statistical ensemble, which does not give rise to 
$\alpha$-effect.\\
Obviously, several generalizations of the model are possible.
In Ref. \cite{S1} Schekochihin \emph{et al.} describe the case 
of a $d$-dimensional smooth velocity field with a generic degree of 
compressibility.
The $\alpha$-effect is taken into account in Refs. \cite{Rog,Boldyrev}
respectively in the limit of large and small Prandtl numbers.
Furthermore, the effect of the Lorenz force is considered 
in Refs. \cite{Boldyrev,S2} 
in the limiting case
of very large Prandtl numbers.

The analysis of dynamo effect is made easier by 
a simple quantum mechanics formulation, first introduced by 
Kazantsev \cite{Kazantsev}.
Indeed, on account of
the $\delta$-correlation in time of the
Kraichnan velocity field, the single time correlation
function for the magnetic field $\left< B_i ( {\bf x} , t )
B_i({\bf x}+{\bf r},t)\right>$ can be expressed in terms
of a function that satisfies a one-dimensional Schr\"odinger-like equation.
The problem of the dynamo effect can thus be mapped into that 
of studying the bound states of a quantum particle in a 
given potential that only depends on the velocity correlation function.
In particular, the ground state energy $E_0$ of such potential
will turn out to be the asymptotic
magnetic field growth rate.
The technique used to compute $E_0$ for different quantum potentials
is the variation-iteration method described in appendix \ref{app:B}.

The aim
of this paper
is to single out the role that the velocity scales $L$ and $\eta$
play in dynamo theory.
To this purpose it is interesting to study the magnetic field generation
as the magnetic Reynolds number $Re_m$ and the Prandtl number $Pr=\nu/\kappa$
are varied ($\nu$ denotes the viscosity of the fluid). Indeed, 
they are related to the relative importance of the 
characteristic scales in the physical problem by the expressions
$Re\simeq L/r_d$ and $Pr\simeq (\eta/r_d)^\xi$.\\
In Ref. \cite{Kazantsev} Kazantsev finally restricts himself to 
the limiting case of $Re_m\to\infty$ and $Pr\to 0$ and proves that 
dynamo can take place only for a velocity scaling exponent 
in the range $1\leq \xi\leq 2$.
Here we show
that in this range of $\xi$ 
the characteristic time of the dynamo effect is of order of
the diffusive time $t_d=r_d^2/\kappa=O(|{\mathbf B}|/|\kappa
\nabla^2{\mathbf B}|)$ and the magnetic field correlation length is of order
$r_d$. We also provide a numerical computation of the growth 
rate vs $\xi$.\\
Then, we analyze the case of finite $Re_m$ and prove that a critical 
magnetic Reynolds number exists: if $Re_m$ is sufficiently small, the 
dynamo does not ever take place, even for $1\leq\xi \leq2$.\\
Finally, we show that the Prandtl number does not affect the presence
of dynamo, but only determines the magnetic field correlation length
and the characteristic growth time. If $Pr<1$, the field $\mathbf{B}$
has a correlation length of order $r_d$ and it grows with a characteristic
time scale of order $t_d$. On the contrary, if $Pr>1$, the correlation
length is of order $\eta$ and the characteristic time of order $t_v$, where
$t_v$ represents the viscous time for the velocity field: 
$t_v=\eta^2/\nu=O(|{\mathbf v}|/|\nu
\nabla^2{\mathbf v}|)$.

This paper is organized as follows. After this
general introduction to the problem, in section \ref{K-K} we define
more precisely the Kraichnan model 
and, following Kazantsev \cite{Kazantsev}, we describe the quantum formalism
mentioned above. In particular we give the Schr\"odinger equation
that is the central point of this quantum approach.
In section \ref{sec:T-D} we revisit the case of infinite 
magnetic Reynolds number and zero Prandtl number. Then, starting
from these results, we analyse the effect of finite $Re_m$ and we 
study how a nonzero $Pr$ influences the magnetic field evolution in time.

\section{The Kraichnan-Kazantsev model}
\label{K-K}

In this section we recall in detail the quantum formalism
introduced by Kazantsev in Ref. \cite{Kazantsev}.
The random velocity field is assumed to be incompressible, Gaussian,
homogeneous, isotropic, parity invariant,
and $\delta$-correlated in time.
Under these hypotheses it is completely defined by its
correlation matrix
\begin{equation}
\begin{split}
\left< v_i({\bf x},t)v_j({\bf x'},t)\right>&=
\delta(t-t'){\mathcal D}_{ij}({\bf r})\\
&=\delta(t-t')[{\mathcal D}_{ij}(0)-S_{ij}({\bf r})] 
\qquad ({\bf r}={\bf x}-{\bf x'}),
\end{split}
\end{equation}
where $S_{ij}(\mathbf{r})$ denotes the structure 
function of the field $\mathbf{v}$.\\
The $\delta$-correlation in time of  $\mathbf{v}$ 
is an essential property in order to write 
a closed equation for the magnetic field correlation function
that, under a suitable 
transformation, reduces to a Schr\"odinger-like equation.

We impose homogeneous and isotropic 
initial conditions for $\mathbf{B}$. Therefore,
on account of the translational and rotational
invariance of Eqs. \eqref{B},
the magnetic field maintains homogeneous and isotropic statistics
at every time $t$.
Its correlation tensor has thus the form (see, e. g., Ref. \cite{Monin})
\begin{equation}
\left< B_i ({\bf x},t) 
B_j({\bf x}',t)  
\right>=G_1(r,t)\delta_{ij}+G_2(r,t)
\dfrac{r_i r_j}{r^2}.
\end{equation}
Because of the solenoidality condition $\nabla\cdot{\mathbf B}
=0$, the functions
$G_1$ and $G_2$ are related by the following differential
equation
\begin{equation}
\dfrac{\partial G_1}{\partial r}=
-\dfrac{1}{r^2}\dfrac{\partial}{\partial r}(G_2 r^2).
\end{equation}
The covariance of $\mathbf{B}$ is then completely described
by a single scalar function, e. g., its trace $H(r,t)=
3\,G_1(r,t)+G_2(r,t)$. The dynamo effect will  correspond
to an unbounded growth in time of $H(r,t)$.\\
The correlation function $H(r,t)$ can be transformed into another
function $\Psi(r,t)$ that
solves
the imaginary time Schr\"odinger equation
\begin{equation}\label{Schr-t}
-\dfrac{\partial\Psi}{\partial t}+ \left[ \dfrac{1}{m(r)}
\dfrac{\partial^2}{\partial r^2}
-U(r)\right]\Psi=0
\end{equation}
in which the mass and the potential depend on $r$ only
through $S_{ii}(r)$. (For the details see the appendix \ref{app:A}
and Ref. \cite{Kazantsev}).\\
To study the dynamo effect it is 
useful to put in evidence the time dependence of $\Psi$.
As usual in quantum mechanics, we thus expand the `wave
function' $\Psi$ in terms of the `energy' eigenfunctions
$\Psi(r,t)=\int \psi_E(r)e^{-Et}\varrho(E)dE$
(or $\Psi(r,t)=\sum_E \psi_E(r)e^{-Et}$ for discrete energy levels)
and  get the `stationary' equation
\begin{equation}\label{staz}
\dfrac{1}{m(r)}\dfrac{d^2 \psi_E}{dr^2}
+[E-U(r)]\psi_E=0.
\end{equation}
Referring to the meaning of $\Psi$, it is clear that an 
unbounded growth of the magnetic field corresponds to the
existence of negative energies in Eq. \eqref{staz}.
In particular, it is the sign of the ground state energy
$E_0$ that determines the presence of dynamo and its value
eventually represents the asymptotic growth rate of the magnetic field.
Indeed, in this case
it is the ground state $\psi_{E_0}e^{-E_0t}$ that dominates 
the growth in time. (Recall that the negative energy levels of a
Schr\"odinger equation are always discrete).

By looking at the variational expression for the
eigenvalues in \eqref{staz}
\begin{equation}\label{variaz}
E=\dfrac{\int mU\psi_E^2 dr +\int (\psi'_E)^2 dr}
{\int m\psi_E^2 dr},\end{equation}
one can easily conclude that the presence
of dynamo effect is equivalent to the existence
of bound states for a quantum particle of
unit ($r$-independent) mass 
in the potential $V(r)=m(r)U(r)$ \cite{V96}.
Therefore, 
in order to
state if dynamo can take place for a given velocity field,
it is sufficient to study the properties of $V$.

Having summarized
the quantum
mechanics formalism
for a magnetic field transported by a Kraichnan turbulent
flow, in the next section we study the dynamo effect for
a velocity correlation function
that mimics the real physical situation.
In particular, we numerically compute $E_0$ and describe 
the properties of the ground state eigenfunction as $Re_m$ 
and $Pr$ are varied. From this analysis we are able to obtain
information about the critical magnetic Reynolds number, 
the correlation length of the magnetic field, 
the asymptotic behaviours of its correlation function,
and the characteristic time-scale of the magnetic field growth.

\section{Turbulent dynamo}
\label{sec:T-D}
We consider the 
realistic situation of a structure function $S_{ii}(r)$
that scales as $r^2$ for $r\ll \eta$, 
as expected in the viscous range,
as $r^\xi$
$(0\leq \xi \le 2)$ in the inertial range 
$\eta\ll r\ll L$, and tends to a constant value ${\mathcal D}_{ii}(0)$
for $r\gg L$.\\
The case $\xi=0$ corresponds to the diffusive behaviour, while the other 
limit $\xi=2$
describes a velocity field that is smooth at all scales below the 
correlation length $L$.
For the other values of  
$\xi$, in the inertial range $S_{ii}(r)$
is only an H\"older continuous function of $r$ with exponent
$\xi/2$. The parameter $\xi$ thus represents a measure of the field
roughness. 

An explicit expression for the velocity correlation tensor, 
which has the desired scaling properties, is, for example,
\begin{gather}
{\mathcal D}_{ij}({\mathbf r})=\int e^{i {\mathbf k}\cdot{\mathbf r}}
\widehat{{\mathcal D}}_{ij}({\mathbf k})
d^3{\mathbf k}\\
\intertext{with}
\label{Dk}
\widehat{{\mathcal D}}_{ij}({\mathbf k})=
D_0\dfrac{
e^{-\eta k}}{(
k^2+L^{-2})^{(\xi+3)/2}}
P_{ij}({\mathbf k}).
\end{gather}
The solenoidal projector $P_{ij}({\mathbf k})=
(\delta_{ij}-k_ik_j/k^2)$
ensures the incompressibility of the velocity field.\\
In what follows we refer to Eq. \eqref{Dk} whenever we show
numerical computations that exemplify our conclusions. 
However, it should be noted that our results are general:
they depend only on the qualitative properties of $S_{ii}(r)$ and
not on its explicit form.

\subsection{Fully developed turbulent dynamo}

We first consider the limiting case of $Re_m\to \infty$ and $Pr\to~0$.
Under these conditions the diffusive scale $r_d$ is in the inertial range
and the presence of the cutoffs $L$
and $\eta$ is neglected: 
only the scaling behaviour $r^\xi$ $(0\leq \xi\leq 2)$ is
considered for the velocity structure function.\\
The general expression of $S_{ij}(r)$ for an homogeneous,
isotropic, parity invariant, incompressible field that scales as $r^\xi$
is \cite{Monin}
\begin{equation}
\underset{L\to \infty}{\lim_{\eta\to 0}} 
S_{ij}({\mathbf r})=
D_1r^{\xi}\,\left[(2+\xi)\delta_{ij}-
\xi\dfrac{r_i r_j}{r^2}\right]
\end{equation}
where the coefficient $D_1$ has the dimensions of $length^{(2-\xi)}/time$.\\
In this limit the total energy ${\mathcal D}_{ii}(0)$
diverges with the infrared cutoff as $L^\xi$.

In order to analyze the existence of the dynamo, 
let us turn to the quantum formulation
described above.
The potential $V$ has the following asymptotic
behaviours (see the appendix \ref{app:A} and Ref. 
\cite{Kazantsev}
for the complete expression)
\begin{equation}
\label{asint1}
V(r)\sim
\begin{cases}
2/r^2 &r\ll r_d\\
(2-\frac{3}{2}\xi-\frac{3}{4}\xi^2)/r^2& r_d\ll r.
\end{cases}
\end{equation}
For sufficiently small $\xi$ the potential is positive for all
$r$, it does not generate bound states and therefore the dynamo 
cannot take place. For larger $\xi$, $V$ is repulsive
up to $r\simeq r_d$ and becomes attractive at infinity (Fig. 
\ref{fig:1}).
A quantum mechanical analysis based on asymptotic behaviours
\eqref{asint1} allows to establish that $\xi=1$
is the exact threshold for the dynamo effect \cite{V96}.

If $0\leq \xi\leq 1$ the turbulent flow alone is unable to 
increase the magnetic field and $B^2$ decays in time.
For those values of $\xi$, the presence of a forcing
term in Eq. \eqref{B} is necessary to obtain a statistically 
stationary state. See Vergassola \cite{V96} for the detailed analysis.

\begin{center}
\begin{figure}
\psfig{file=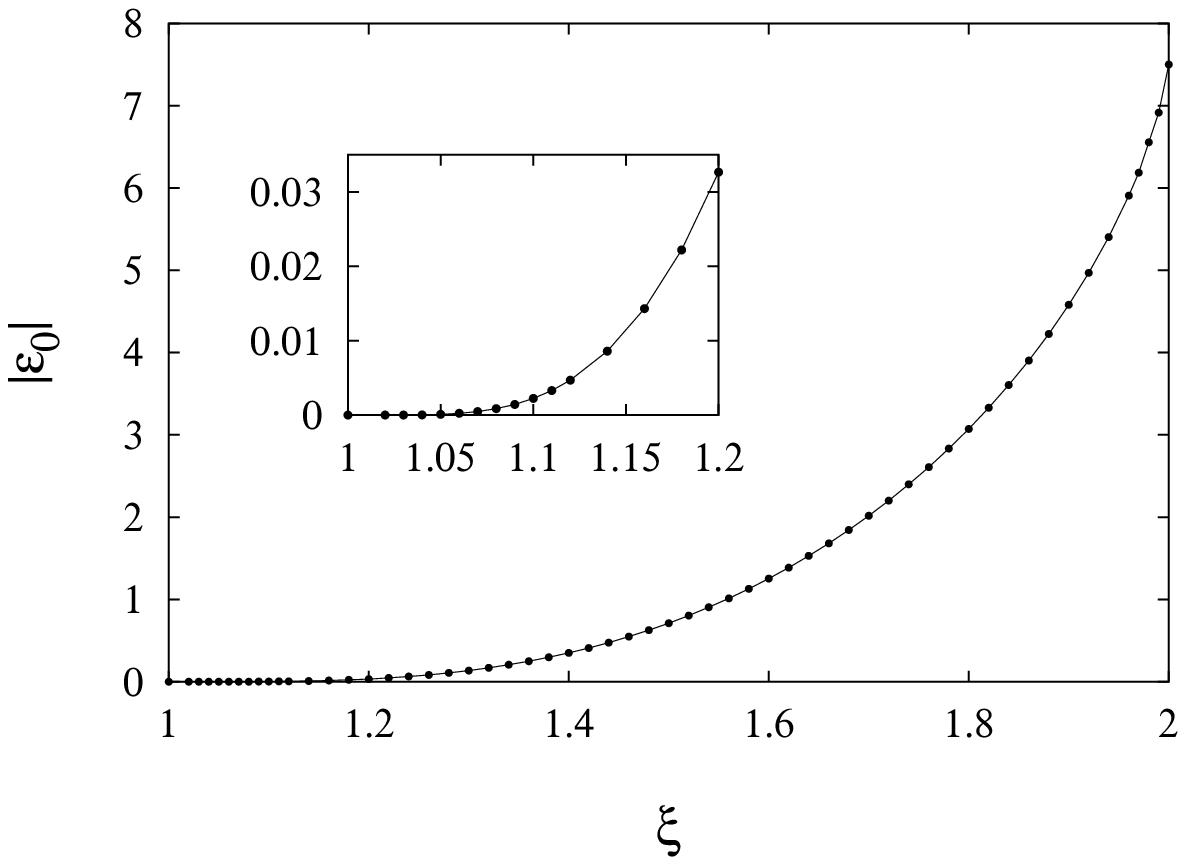}
\vspace{1cm}
\caption{D. Vincenzi}
\label{fig:2}
\end{figure}
\end{center}

\begin{center}
\begin{figure}
\psfig{file=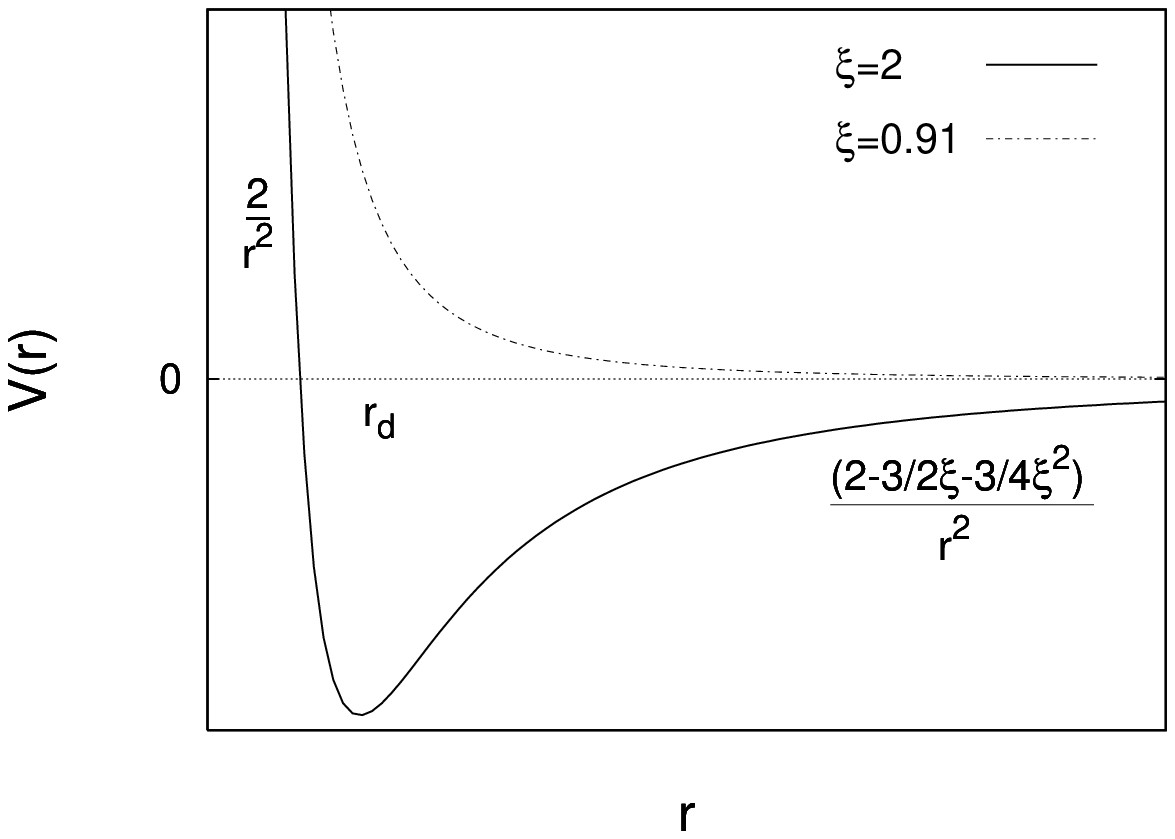}
\vspace{1cm}
\caption{D. Vincenzi}
\label{fig:1}
\end{figure}
\end{center}

From now on we restrict to the values $1\leq\xi\leq 2$, for
which the dynamo is present.\\
If Eq. \eqref{staz} is rewritten in a rescaled form
by means of the transformation $r\rightarrow r/r_d$, $r_d=(\kappa /D)^{1/\xi}$,
it is easy to see that the eigenvalues of the energy must take the
form 
\begin{equation}\label{time}
E=\epsilon(\xi)t_d^{-1},\end{equation} 
where $\epsilon(\xi)$ depends only on the 
scaling exponent $\xi$ and $t_d$ 
is the characteristic time of  
magnetic diffusion.\\
We have already noted that the ground state eigenfunction dominates the
evolution in time and that $E_0$
is the asymptotic
magnetic growth rate. 
We numerically compute $\epsilon_0(\xi)$ as a function of 
$\xi$ by the variation-iteration method described in the 
appendix \ref{app:B}. The quantity $\epsilon_0$ grows with $\xi$ as shown in 
Fig. \ref{fig:2}. 
When $\xi$ tends to one, $\epsilon_0$ approaches zero and the
bound states disappear. In the other limit, $\epsilon_0$ reaches
the value $15/2$ according to the results of Kazantsev \cite{Kazantsev}.
An estimation for $\epsilon_0$ vs $\xi$ already appears 
in Ref. \cite{Kazantsev},
but there the results are limited to the values $1.25<\xi<2$. Moreover,
the numerical computations in that paper are performed by a
variational method based on the particular guess $r^2e^{-\beta r}$
for the eigenfunction $\psi_{E_0}$. 
This ansatz is correct for $r\ll r_d$, but it fails to capture the right
behaviour for $r\gg r_d$.
Indeed, if we 
insert the asymptotic behaviours 
\eqref{asint1} in Eq. \eqref{staz}, we find that, for $1<\xi<2$, 
$\psi_E(r)$ shows for $r\gg r_d$ a stretched exponential decay with
characteristic scale $r_d$ and stretching exponent $(2-\xi)/2$
(Fig. \ref{fig:3}).
The variation-iteration method we used (see the appendix \ref{app:B})
presents the big advantage of not
requiring an explicit form for $\psi_{E_0}$. The algorithm provides as 
results both the eigenvalue and the corresponding eigenfunction.

From the expressions of $\psi_{E_0}(r)$ 
we can recover the behaviour of $H(r,t)$
(see the definition \eqref{Psi} in
appendix \ref{app:A}).
We have that, for $r\ll r_d$, $H(r,t)$ is approximately constant, while,
if $1<\xi<2$, 
the magnetic field correlation function decays for $r\gg r_d$ as a
stretched exponential with scale $r_d$
\begin{equation}
H(r,t)\propto e^{-\beta\,\left( r/r_d\right)^{(2-\xi)/2}}
\qquad (r_d\ll r\ll L).
\end{equation}
The prefactor
\begin{equation}
\beta=
\dfrac{\sqrt{2|\epsilon_0(\xi)|}}{2-\xi}
\end{equation}
depends on the growth rate $\epsilon_0(\xi)$. 
We can thus conclude that,
for $1<\xi<2$, the magnetic field has a spatial distribution characterized
by structures whose  scales are of order $r_d$.\\
The cases $\xi=2$ and $\xi=1$ 
have to be treated separately. Indeed, the asymptotic 
properties cannot be deduced directly from Eq. \eqref{staz}.\\
The smooth case is solved by Chertkov \emph{et al.} in Ref. \cite{V99}
by a Lagrangian approach that relates the growth rate to the 
Lyapunov exponents. 
There is a big difference between the situation 
of a smooth velocity field and
one that is just H\"older continuous. 
In the former case
the correlation function is found to depend on the spatial
coordinate as $H(r,t)\propto r^{-5/2}$
(equivalent to $\psi_{E_0}(r)\propto r^{1/2}$),
which implies the presence of structures with at least one dimension
of inertial range size. Actually the magnetic field in the smooth case
has been shown to be characterized by strip-like objects.\\
Finally, the case $\xi=1$ can be solved exactly. Indeed, the
appropriate ground state eigenfunction of 
Eq. \eqref{staz} is 
(recall that $\xi=1$ is the threshold for dynamo and 
hence $E_0=0$)
\begin{equation}
\psi_{0}(x)={C\,\frac{\,{\sqrt{1 + x}}\,
     \left( -2\,x + 
       \left( 2 + x \right) \,\log (1 + x)
        \right) }{x}}, \qquad (x=r/r_d), 
\end{equation}
where
the constant $C$ is related to the value of $H(0,\cdot)$ by the 
relation $C=3\sqrt{\kappa}\,r_d^2\, H(0,\cdot)$.\\
If we neglect logarithmic corrections, the asymptotic 
behaviour of $\psi_0$ for $r\gg r_d$ is $\psi_0(r)\propto~
r^{1/2}$, which yields again $H(r,\cdot)\propto~ r^{-5/2}$
for $r\gg r_d$.
                               
The results we have outlined in this section will be useful in the
following to describe the general case 
where the velocity energy spectrum has an infrared and an ultraviolet
cutoff. Indeed, we will study a structure
function that for $r\ll\eta$ scales as $r^2$ and so takes the $\xi=2$
behaviour, while for $r\gg L$ tends to a constant value like in the diffusive 
case $\xi=0$.

\subsection{Finite Reynolds effect}
Let us analyze the situation of finite $Re_m$
(and zero Prandtl number). The principal fact is that a large scale
cutoff $L$ appears for velocity field correlations.
The diffusive scale $r_d$ is again
within the inertial range of the velocity fluctuations,
and the presence of the viscous cutoff can be neglected.
The velocity structure function therefore
scales as $r^\xi$ for $r\ll L$ and tends
to ${\mathcal D}_{ii}(0)$ for $r\gg L$.\\
Therefore, the potential $V$ behaves as in the previous case for 
$r\ll L$, while it takes the $\xi=0$ behaviour for $r\gg L$
\begin{equation}
V(r)\sim
\begin{cases}
2/r^2 & r\ll r_d\\
(2-\frac{3}{2}\xi-\frac{3}{4}\xi^2)/r^2&r_d\ll r\ll L\\
2/r^2 & L\ll r.
\end{cases}
\end{equation}
The main consequence of a finite $Re_m$ 
is that $V$ is repulsive also at large scales. It is thus clear
that, for sufficiently high $Re_m$, a potential well
is present at scales of order $r_d$. On the contrary, if $Re_m$ is too
small, the well can be absent or anyway not deep enough
to generate bound states (see Fig. \ref{fig:4}). Therefore,
we can conclude that for sufficiently small
$Re_m$, the dynamo does not take place, even for $1<\xi<2$.\\
The effect of a large scale cutoff on the velocity energy spectrum 
is thus the presence of a critical Reynolds number $Re_m^{(crit)}$.
For $Re_m$ smaller than that value the potential $V$ 
has not bound states or equivalently, 
on account of our quantum mechanic interpretation,
the velocity field is unable to favour the magnetic field growth and the
ohmic dissipation eventually prevails on stretching.\\
The dependence of 
the dimensionless rate-of-growth
$E_0\,t_d^{-1}$ on $Re_m$ 
is shown in Fig. \ref{fig:5}
for $Re_m>Re_m^{(crit)}$ in the case of the scaling $\xi=4/3$.
Notice that, for $Re_m\gg Re_m^{(crit)}$, $E_0$ takes the inertial
range behaviour $E_0\simeq\epsilon_0(\xi)t_d^{-1}$. 

We can again deduce from Eq. \eqref{staz} some properties
of the function $H(r,t)$. The correlation length of the magnetic 
field is again of order $r_d$ and, at $r\gg L$, $H(r,t)$ shows an
exponential decay
\begin{equation}
H(r,t)\propto
e^{-\gamma\,(r/L)}             
\qquad (L\ll r),
\end{equation}
with $\gamma=E_0\,[L^2/(2\bar{\kappa})]^{-1}$,
$\bar{\kappa}=\kappa+{\mathcal D}_{ii}(0)/6$.

\subsection{Nonzero Prandtl effect}
Finally, we consider the situation of nonzero Prandl number (at infinite 
Reynolds  number). This is equivalent to look at the effect of the viscous 
scale on the dynamo effect.\\
If $Pr<1 $, the diffusive scale $r_d$ is in the inertial range, while, if
$Pr>1$, it lies within the viscous range.
The structure function $S_{ii}(r)$ scales as $r^2$ for $r\ll\eta$ and
as $r^\xi$ for $r\gg\eta$.\\
From the previous considerations we can expect 
for the potential $V$ the same asymptotic behaviours for $r\to\infty$ 
as in 
the case of $Pr=0$. Therefore, the Prandtl number does not influence
the presence of dynamo.
What is sensitive to $Pr$ is the correlation length of the magnetic
field, that approximately corresponds to the scale at which
the function $\psi_{E_0}$ begin its exponential-like decay.
When $Pr<1$, the potential has nearly the same shape as in the case
$Pr=0$
\begin{equation}
V(r)\sim
\begin{cases}
2/r^2 & r\ll r_d\\
(2-\frac{3}{2}\xi-\frac{3}{4}\xi^2)/r^2 & r_d\ll r
\end{cases}
\end{equation}
and the correlation length is of order $r_d$.\\
On the contrary, when $Pr>1$, 
the potential well is modified by an attractive $\xi=2$ contribution
\begin{equation}
V(r)\sim
\begin{cases}
2/r^2 & r\ll r_d\\
-4/r^2 & r_d\ll r\ll \eta\\ 
(2-\frac{3}{2}\xi-\frac{3}{4}\xi^2)/r^2 & \eta\ll r.
\end{cases}
\end{equation}
For these $Pr$ the function $\psi_{E_0}(r)$ grows as $r^2$ for
$r\ll r_d$, as $r^{1/2}$ in the range $ r_d\ll r\ll \eta$ and 
has a stretched exponential decay for $\eta\ll r$.  
We can thus conclude that, when $Pr>1$, the magnetic field correlation
length is of order $\eta$.

On account of what we have just seen, we expect that for $Pr\ll 1$ the ground 
state energy will be proportional to the diffusive time:
$E_0\simeq \epsilon_0(\xi) t_d^{-1}$.
In the other limit, $Pr\gg 1$, 
we can predict an approximate expression for $E_0$
by a simple scaling argument. Indeed, for large $Pr$ the potential $V$
behaves like  in the case $\xi=2$ and we can expect $E_0\propto D_1$
(see Ref. \cite{V99} for the discussion of the smooth case).
Knowing that $S_{ii}(r)\propto r^2$ for $r\ll \eta$ and
$S_{ii}(r)\propto D_1r^\xi$ in the inertial range,
we can match the previous behaviours to obtain $D_1\propto \eta^{\xi-2}$.
Finally, we
recall that from dimensional arguments we have $\eta\simeq
(\nu/D_1)^{1/\xi}$. Summarizing the previous considerations,  
it is easily seen that, for $Pr\gg 1$, the relation $E_0\propto
t_v^{-1}$ holds (the time $t_v$
 is the characteristic one for the 
velocity diffusion).\\
The Prandtl number $Pr\simeq(t_v/t_d)^{\xi/(\xi-2)}$
thus influences also the magnetic field rate-of-growth:
in presence of dynamo, $B^2$ increases with a
characteristic time-scale determined by the largest beetween the viscous
and the diffusive time (see Fig. \ref{fig:6}).

To conclude this section, we discuss a result that emerges from numerical
computations: for $Pr\simeq 1$ the magnetic growth rate reaches a
maximum (Fig. \ref{fig:6}). 
We can easily guess this behaviour, if we refer once more
to the Kazantsev quantum formalism.
For $Pr<1$ the $\xi=2$ behaviour is practically absent in the potential
$V$, while, when $Pr$ approaches the value $1$, the scale $\eta$ 
begins to come into play yielding a 
strongly attractive $-4/r^2$ contribution at scales
$r_d\ll r\ll \eta$. The $\xi=2$ potential is more attractive than that 
of other $\xi$ and the ground state energy increases in 
absolute value. Then, as $Pr$ becomes larger, $|E_0|$ decreases
as explained above. In other words as long as the viscous
behaviour affects only the potential shape around $r_d$, its only
effect is to make the well deeper and so to favour the dynamo.
When viscousity becomes very large, the level of velocity 
fluctuations lowers significantly, inducing eventually the depletion
of the rate-of-growth.

\section{Conclusions}
The presence of dynamo effect for a magnetic field advected
by a conducting fluid is strongly dependent on the 
properties of the turbulent flow. We have highlighted in the 
framework of the Kraichnan ensemble the consequences of
considering a viscous and an integral cutoff for the velocity.
Using the quantum formalism introduced by Kazantsev, we have
found that a critical magnetic Reynolds number exists.
By the same method, we have shown that the Prandtl number 
is the parameter that determines the correlation length 
of the magnetic field and the characteristic time of its growth.
Finally, we have argued that in the presence of dynamo the magnetic growth rate
is maximum fo Prandtl number of order unity. 
As already noted, the previous analysis depends only on the qualitative
properties of the velocity structure function, so we expect our conclusions
to hold for a generic turbulent flow with same statistical 
simmetries and therefore to be relevant for real applications.

\acknowledgments

I would like to express my gratitude to A. Celani for his
fundamental contribution in developing this work. 
A. Mazzino and M. Vergassola are
also acknowledged for useful discussions and suggestions. This work
was supported by the doctoral grants of the Nice University,
by the Department of Physics of the Genova University, and by the
European Union under Contract HPRN-CT-2000-00162.

\appendix

\section{Variation-Iteration Method}
\label{app:B}
For the numerical analysis of Schr\"odinger equation \eqref{staz}
we make the transformation $y=a^{-r}$ $(a>1)$ which maps $(0,\infty)$ 
on the finite interval $(0,1)$. (The constant $a$ should be chosen
to properly resolve this interval).
Eq.\eqref{staz} can thus be rewritten in 
the form
\begin{gather}
 \label{op.}
  {\mathcal L}\,\psi = \lambda\, {\mathcal M}\, \psi\\
\intertext{where}
\begin{array}{c}
{\mathcal L}=-(\ln a)^2\left(y\dfrac{d^2}{dy^2}
+\dfrac{d}{dy}\right)+\dfrac{m(y)}{y}\left(U(y)-U_{min}
\right)\\[0.5cm]
{\mathcal M}=\dfrac{m(y)}{y}\qquad\qquad
\lambda=E-U_{min}\end{array}
\end{gather}
and $U_{min}$ denotes the minimum value of $U$.
${\mathcal L}$ and ${\mathcal M}$ are positive-definite self-adjoint operators
defining a spectrum of eigenvalues $\lambda$ bounded from below
and which extends to infinity.
Moreover, 
${\mathcal L}$ is invertible on all functions twice differentiable on $(0,1)$
and vanishing at the boundaries of the interval.
Under these hypotheses the variation-iteration method described
in Ref. \cite{Morse} provides a valuable tool to compute
the lowest eigenvalue
$\lambda_0$ of Eq. \eqref{op.}
and the corresponding eigenfunction $\psi_0$. Indeed,
let $\varphi_0$ be an initial trial function such that $\int_0^1\psi_0\,
{\mathcal M} \varphi_0\, dy \ne 0$ and define the $n$th iterate $\varphi_n$ as 
\begin{equation}
\varphi_n\equiv {\mathcal L}^{-1}
{\mathcal M}\varphi_{n-1}=({\mathcal L}^{-1}
{\mathcal M})^n\varphi_0.
\end{equation} 
Then, as $n$ is increased, the sequence $\varphi_n$ converges to the
eigenfunction $\varphi_0$.
The $n$th approximation to $\lambda_0$ is given by the following 
variational expression employing $\varphi_n$ as trial function
\begin{equation}
\lambda_0^{(n)}=\dfrac{\displaystyle \int_0^1\varphi_n 
{\mathcal L} \varphi_n dy}{
\displaystyle \int_0^1\varphi_n {\mathcal M} \varphi_n dy}.
\end{equation}
The set $\lambda_0^{(n)}$ form a monotonic sequence
of decreasing values, approaching $\lambda_0$ from the above.
The advantage of the variation-iteration technique
is that no expression is required
\emph{a priori} for the function $\psi_0$. We only have to choose any 
guess for  initial function $\varphi_0$ and then improve the result 
by iterating the method for sufficiently large $n$. 
The convergence is more rapid the smaller
is the ratio between $\lambda_0$ and the following eigenvalue.

Finally, for the numerical implementation of the method, 
we exploited the first order
discrete expression of ${\mathcal L}$ 
preserving the boundary conditions on $\psi$.
If $(0,1)$ is divided in intervals of
length $\Delta$ and $y_i=i\Delta$, we have
\begin{equation}
{\mathcal L}_{ij}=
\dfrac{m(y_i)}{y_i}(U(y_i)-U_{min})
+\dfrac{(\ln a)^2}{2 \Delta^2}\times
\begin{cases}
 (-y_{i-1}-y_i)&\text{if $i=j+1$}\\
 (y_{i-1}+2y_i+y_{i+1})&\text{if $i=j$}\\
 (-y_i-y_{i+1})&\text{if $i=j-1.$}
\end{cases}
\end{equation}

\section{The Schr\"odinger equation in the dynamo theory}
\label{app:A}
In the present appendix we refer to the notation adopted in the 
body of the paper. So, the trace of the correlation tensor 
$\left<B_i({\mathbf x},t)B_j({\mathbf x}+{\mathbf r},t)\right>$ will
be denoted by $H(r,t)$.\\
As a consequence of the velocity $\delta$-correlation in time, 
$H$ satisfies a closed equation that, under a 
suitable transformation, takes on the form of a 
one-dimensional Schr\"odinger-like
equation. In order to exploit this fact, let us denote $s(r)=S_{ii}(r)$
and define the following quantities
\begin{equation}
\begin{array}{lcr}
&\overline{s}(r)=\dfrac{1}{r^3}\displaystyle{\int_0^r}
\dfrac{s(\rho)}{2}\rho^2d\rho,&\\
\Lambda(r)=\kappa+\overline{s}(r),&&
\Lambda_1(r)=\Lambda(r)+3\kappa+\dfrac{s(r)}{2}.
\end{array}
\end{equation}
Then, the function
\begin{equation}
\label{Psi}
\Psi(r,t)=\sqrt{\kappa}\,\exp\left(\int_0^r \dfrac{\Lambda_1(\rho)}{
2\rho\Lambda(\rho)}d\rho\right)\,\dfrac{1}{r^3}
\int_0^r H(\rho,t)\rho^2d\rho
\end{equation}
solves the imaginary time Schr\"odinger equation
\begin{equation}
\label{Schr;t}
-\dfrac{\partial\Psi}{\partial t}+ \left[ \dfrac{1}{m(r)}
\dfrac{\partial^2}{\partial r^2}
-U(r)\right]\Psi=0
\end{equation}
where
\begin{equation}
\label{massa}
m=\dfrac{1}{2\Lambda},
\qquad
U=-\dfrac{1}{r}\dfrac{ds}{dr}+\dfrac{1}{2r^2}
\dfrac{\Lambda_1^2}{\Lambda}+\Lambda\dfrac{d}{dr}\left(
\dfrac{\Lambda_1}{r\Lambda}\right) .
\end{equation}
(See Ref. \cite{Kazantsev} for the detailed derivation).
If we expand $\Psi$ in terms of the energy eigenfunctions
$\Psi(r,t)=\int \psi_E(r)e^{-Et}\varrho(E)dE$, we get the stationary
equation
\begin{equation}\label{stationary}
\dfrac{1}{m(r)}\dfrac{d^2 \psi_E}{dr^2}
+[E-U(r)]\psi_E=0.
\end{equation}
The dynamo effect corresponds to the presence of negative 
eigenvalues in Eq. \eqref{stationary}.\\
The correlation function $H(r,\cdot)$ must tends to a constant value 
as $r\to 0$ and decreases to zero as $r\to\infty$. From the definition
\eqref{Psi} we have therefore that Eq. \eqref{stationary} must be
solved with the boundary conditions that $\psi_E(r)$ vanishes as $r\to 0$
and increases as $r\to\infty$ slowly enough to guarantee that 
$H(r,\cdot)$ decreases to zero.
In particular, if $s(r)$ tends to a constant as $r\to\infty$,
$\psi_E(r)$ cannot increase more rapidly than $r$.

We consider now the explicit expression
\begin{equation}
{\mathcal D}_{ij}({\mathbf r})=\int e^{i {\mathbf k}\cdot{\mathbf r}}
\widehat{{\mathcal D}}_{ij}({\mathbf k})
d^3{\mathbf k}
\end{equation}
with
\begin{equation}\label{D}
\widehat{{\mathcal D}}_{ij}({\mathbf k})=
D_0\dfrac{
e^{-\eta k}}{(
k^2+L^{-2})^{(\xi+3)/2}}
P_{ij}({\mathbf k}),\qquad (\alpha>-1).
\end{equation}
The transverse projector $P_{ij}({\mathbf k})=
(\delta_{ij}-k_ik_j/k^2)$
ensures the incompressibility of the velocity field.

In the limits $\eta\to 0$ and $L\to 0$, $S_{ij}(r)$
takes the form
\begin{gather}
\underset{L\to \infty}{\lim_{\eta\to 0}}
S_{ij}({\mathbf r})=
D_1r^{\xi}\,\left[(2+\xi)\delta_{ij}-
\xi\dfrac{r_i r_j}{r^2}\right]\\
\intertext{with}
D_1=\dfrac{4\pi\cos\left(\frac{\pi\xi}{2}\right)
\Gamma(-1-\xi)}{\xi+3}D_0.
\end{gather}
(The function $\Gamma$ is the Euler function).\\
If we insert $s(r)=2(\xi+3)D_1r^\xi$ in \eqref{massa}, 
the transformation \eqref{Psi} takes on the form
\begin{gather}
\Psi(r,t)=\dfrac{(\kappa+D_1r^\xi)^{1/2}}{r}
\int_0^r H(\rho,t)\rho^2d\rho,\\ 
\intertext{while its inverse reads}
H(r,t)=
{\frac{\left( 2\,\kappa - D_1\,{r^{\xi}}\,\left(  
\xi-2 \right)  \right) \,
       \Psi(r,t) + 2\,r\,\left( \kappa 
       + D_1\,{r^{\xi}} \right) \,\Psi'(r,t)}%
{2\,{r^2}\,
      {{\left( \kappa + D_1\,{r^{\xi}} \right) }^{{\frac{3}{2}}}}}}.
\end{gather}
For the mass and the potential
we obtain the following 
expressions 
\begin{gather}
m(r)=\dfrac{1}{2(\kappa+D_1r^\xi)},\\\nonumber\\
U(r)=\dfrac{4\kappa^2+A(\xi)\kappa D_1r^\xi+B(\xi) D_1^2 r^{2\xi}}
{r^2(\kappa+D_1r^\xi)}
\end{gather}
with $A(\xi)=(8-3\xi-\xi^2)$ and $B(\xi)=(4-3\xi-\frac{3}{2}\xi^2)$.\\
For the sake of completeness
we write also the expressions of the trace $s(r)$, 
which we used to compute 
$E_0$ respectively in the case of finite $Re_m$ 
and in the case of nonzero $Pr$
\begin{gather}
\lim_{\eta\to 0}s(r)=
\dfrac{4\pi D_0 L^{\xi}}{\Gamma\left(\frac{\alpha+\xi+1}{2}\right)}
\left[ \Gamma\left(\frac{1+\alpha}{2}\right)\Gamma\left(\frac{\xi}{2}\right)
-\sqrt{\pi}\,\dfrac{L}{r}\,G^{2\ 1}_{1\ 3}\,
\left(\dfrac{r^2}{4L^2}\left|\begin{array}{l}
1-\frac{\alpha}{2}\\
\frac{\xi+1}{2},\frac{1}{2},0
\end{array}
\right.
\right)
\right],
\\\nonumber\\
\lim_{L\to \infty}s(r)=
8\pi D_0\,{{\eta}^{\xi}}
      \left( \Gamma(-\xi) +
       \dfrac{ \eta}{r}
\,{{\left( 1 + {\dfrac{{r^2}}{{{\eta}^2}}} \right) }^
           {{\frac{1 + \xi}{2}}}}\,\Gamma(-1 - \xi)\,
         \sin [\left( 1 + \xi \right) \,\arctan ({\frac{r}{\eta}})] \right). 
\end{gather}
(The function $G$ denotes the $G$-Meijer's function of
argument $r^2/(4L^2)$. See Ref. \cite{G} for the exact definition).
The explicit expressions 
of the mass and the potential
can be derived from \eqref{massa}.


\newpage


\begin{center}
FIGURES CAPTIONS
\end{center}
Fig. 1: The dependence of the magnetic growth rate $\epsilon_0=
E_0 t_d^{-1}$ on the scaling exponent $\xi$ 
in the limit of infinite $Re_m$ and zero $Pr$,
as computed by the 
variation-iteration method described in appendix \ref{app:B}.\\\\
Fig. 2 : The shape of the quantum potential $V$ 
in the limit $Re_m\to\infty$ and $Pr\to 0$ 
for a value of $\xi$ $(\xi=2)$ for which dynamo is present
and for a value $(\xi=0.91)$ for which there is no dynamo effect.
\\ \\
Fig. 3: The asymptotic behaviours of the `stationary wave 
function' $\psi_{E_0}$ in the limit of infinite $Re_m$ and zero 
$Pr$. The maximum at $r\simeq r_d$ determines the 
magnetic field correlation length. \\\\
Fig. 4: 
A qualitative picture of the quantum potential shape
for $Re_m$ respectively above and below the critical value
$Re_m^{(crit)}$ $(1<\xi<2$).
\\\\
Fig. 5: 
The dependence of the magnetic growth rate on the
magnetic Reynolds number for $Pr\to 0$ and $\xi=4/3$.
The numerical computation is performed using expression
\eqref{Dk} for the correlation tensor of the magnetic field.
\\\\
Fig. 6: 
The dependence of the magnetic growth rate on the
Prandtl number for $\xi=4/3$ and in the limit $Re_m\to\infty$.
The numerical computation is performed using expression
\eqref{Dk} for the correlation tensor of the magnetic field.
\\\\

\newpage
\newpage

\newpage

\begin{center}
\begin{figure}
\psfig{file=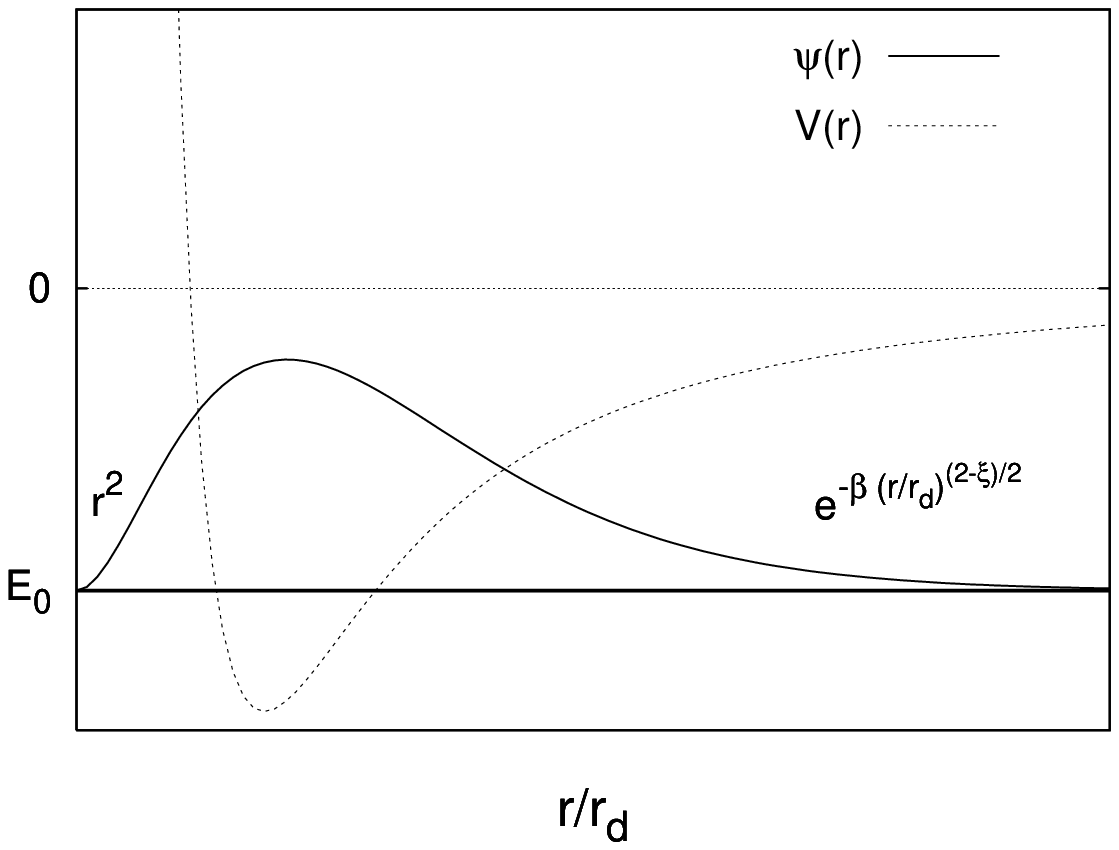}
\vspace{1cm}
\caption{D. Vincenzi}
\label{fig:3}
\end{figure}
\end{center} 
\newpage

\begin{center}
\begin{figure}
\psfig{file=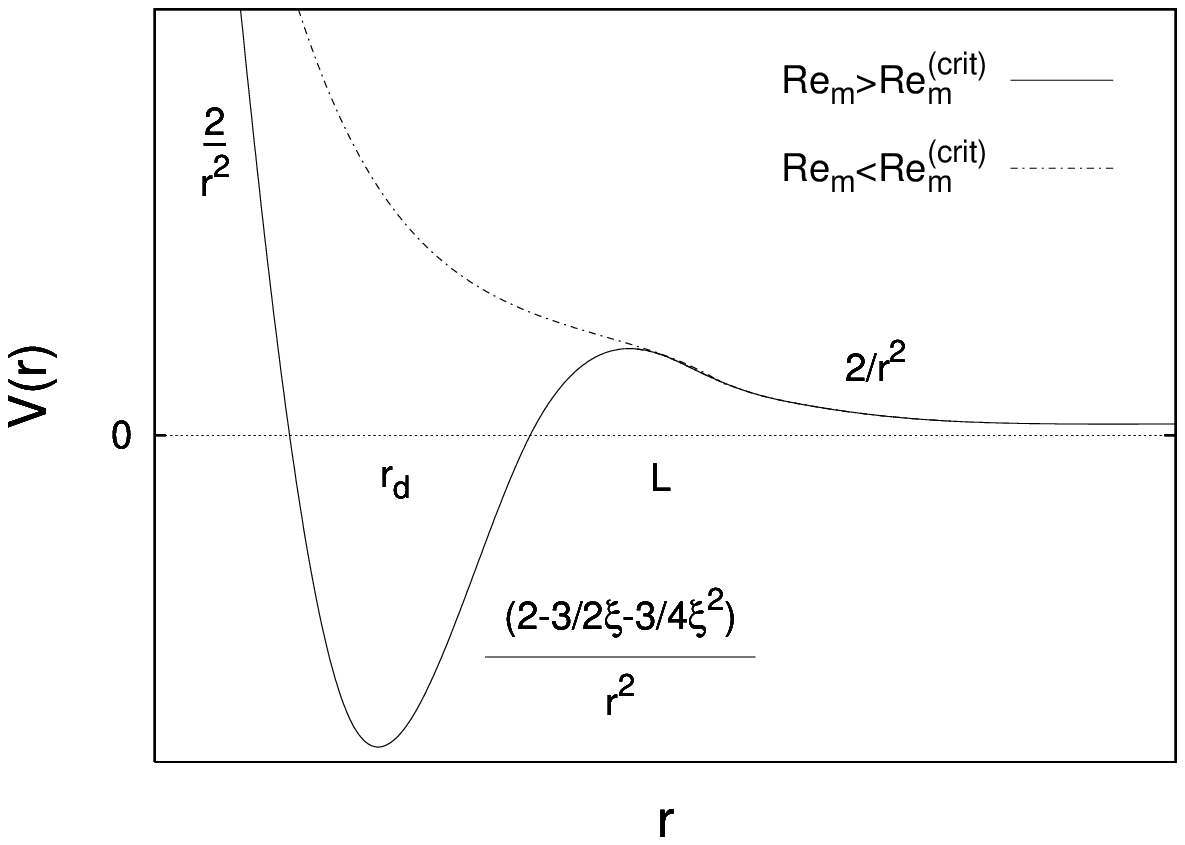}
\vspace{1cm}
\caption{D. Vincenzi}
\label{fig:4}
\end{figure}
\end{center} 
\newpage

\begin{center}
\begin{figure}
\psfig{file=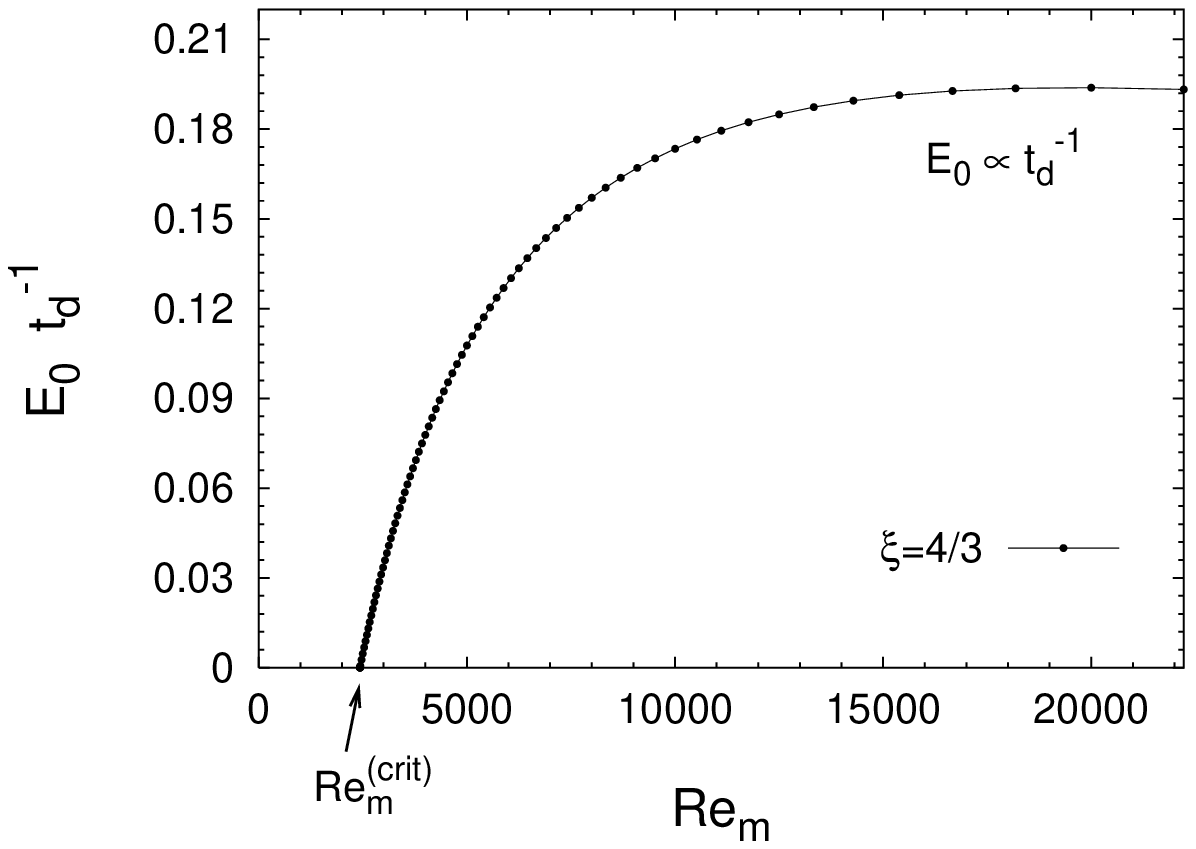}
\vspace{1cm}
\caption{D. Vincenzi}
\label{fig:5}
\end{figure}
\end{center} 
\newpage

\begin{center}
\begin{figure}
\psfig{file=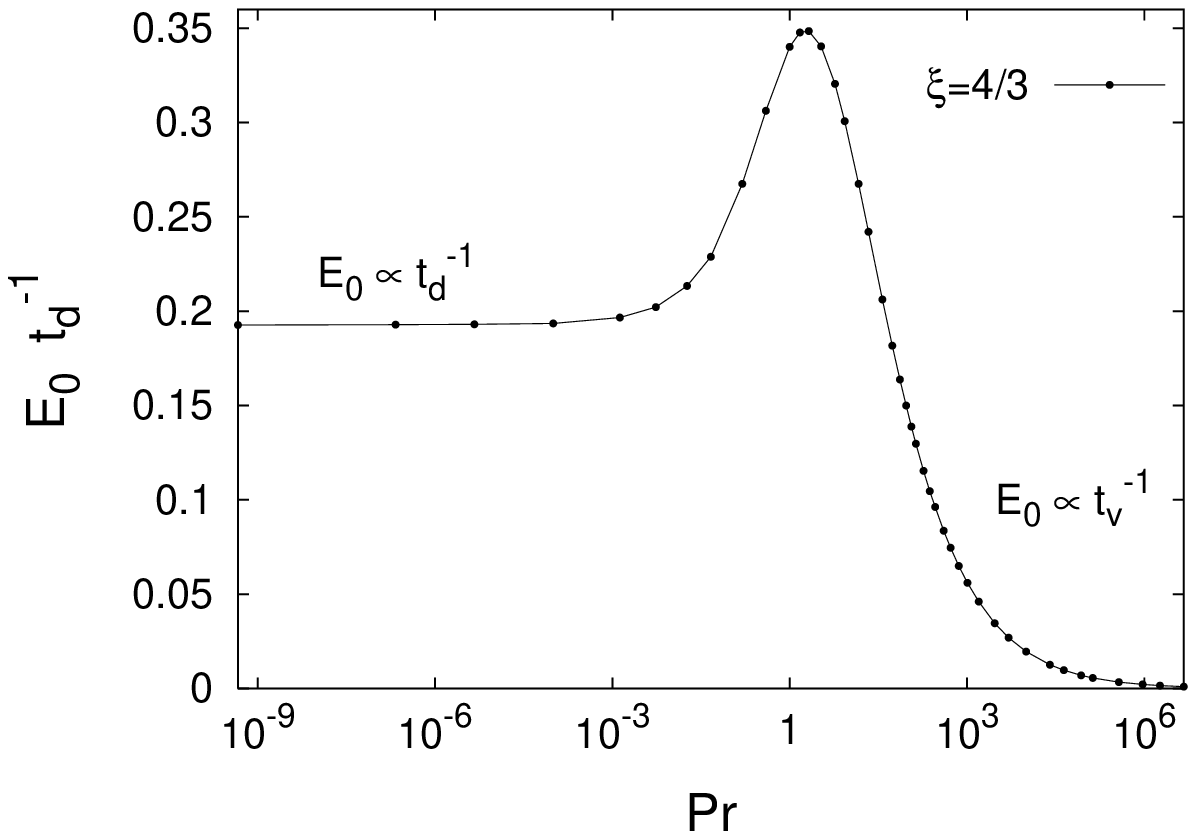}
\vspace{1cm}
\caption{D. Vincenzi}
\label{fig:6}
\end{figure}
\end{center} 

\end{document}